\documentclass[journal,10pt]{IEEEtran}

\usepackage{algorithm,algorithmic}
\usepackage{amsmath,amssymb,mathrsfs}
\usepackage{array}
\usepackage{easyReview}
\usepackage{fancyhdr}
\usepackage{float}
\usepackage{glossaries}
\usepackage{graphicx,color}
\usepackage{hyperref}
\usepackage{psfrag}
\usepackage{stfloats}
\usepackage{subfigure}
\usepackage{url}
\usepackage{booktabs}
\usepackage{bm}
\usepackage{bbm}
\usepackage{mathtools}

\definecolor{sblue}{RGB}{0,51,120}
\definecolor{sred}{RGB}{139,0,139}
\definecolor{sg}{RGB}{46,139,87}

\makeglossaries

\begin{document}

\title{\LARGE \bf Resi-VidTok: An Efficient and Decomposed Progressive Tokenization Framework for Ultra-Low-Rate and Lightweight Video Transmission}

\author{Zhenyu Liu, Yi Ma, Rahim Tafazolli, and Zhi Ding

\thanks{Z.~Liu, Y. Ma and, R. Tafazolli are with 5GIC \& 6GIC,
Institute for Communication Systems (ICS), University of Surrey, Guildford,
U.K. (emails:\{zhenyu.liu; y.ma; r.tafazolli\}@surrey.ac.uk).}
\thanks{Z. Ding is with the Department of Electrical and Computer Engineering, University of California at Davis, USA (e-mail: zding@ucdavis.edu).} 
}
\maketitle

\begin{abstract}
Real-time transmission of video over wireless networks remains highly challenging, even with advanced deep models, particularly under severe channel conditions such as limited bandwidth and weak connectivity. In this paper, we propose \emph{Resi-VidTok}, a Resilient Tokenization-Enabled framework designed for ultra-low-rate and lightweight video transmission that delivers strong robustness while preserving perceptual and semantic fidelity on commodity digital hardware. By reorganizing spatio--temporal content into a discrete, importance-ordered token stream composed of key tokens and refinement tokens, Resi-VidTok enables progressive encoding, prefix-decodable reconstruction, and graceful quality degradation under constrained channels. A key contribution is a resilient 1D tokenization pipeline for video that integrates differential temporal token coding, explicitly supporting reliable recovery from incomplete token sets using a single shared framewise decoder—without auxiliary temporal extractors or heavy generative models. Furthermore, stride-controlled frame sparsification combined with a lightweight decoder-side interpolator reduces transmission load while maintaining motion continuity. Finally, a channel-adaptive source--channel coding and modulation scheme dynamically allocates rate and protection according to token importance and channel condition, yielding stable quality across adverse SNRs. Evaluation results indicate robust visual and semantic consistency at channel bandwidth ratios (CBR) as low as $4\times10^{-4}$ and real-time reconstruction at $>$30~fps, demonstrating the practicality of Resi-VidTok for energy-efficient, latency-sensitive, and reliability-critical wireless applications.
\end{abstract}


\section{Introduction}

Real-time video transmission over wireless networks remains a significant challenge, especially in environments characterized by limited bandwidth and unreliable connections \cite{deepjscc_v_2022tung}. These conditions commonly arise during natural disasters, emergency situations in remote or rural areas, and mobile scenarios involving vehicles or satellites. Under such constraints, traditional methods combining video codecs (H.264/H.265) with advanced channel codes often fail \cite{backgound2}, resulting in severely distorted videos. Even cutting-edge solutions like Apple's emergency message and Huawei's satellite call are limited to basic textual communication \cite{appleMessages2025} and voice calls, highlighting the difficulty of transmitting richer media content over constrained channels.

Semantic communications (SemCom), empowered by artificial intelligence, offer a promising solution by extracting and transmitting semantic features of raw data, substantially reducing communication overhead \cite{deepjscc_v_2022tung, deepjscc_v_2023wang,deepjscc_v_2025bao}. However, these approaches primarily focus on decreasing transmission cost under high-fidelity recovery constraints, limiting their performance in ultra-low-rate scenarios. Meanwhile, using neural network (NN)-based semantic encoders \cite{deepjscc_v_2022tung, deepjscc_v_2023wang,deepjscc_v_2025bao} also makes it difficult to deploy semantic communication systems on modern digital communication devices. Specifically, NN-based semantic encoders usually output continuously distributed signals that are directly sent into the communication channel without being modulated into discrete constellation symbols. Such an analog transmission approach is difficult to implement in practice due to non-ideal hardware characteristics (e.g., power amplifiers) and compatibility issues with existing digital communication protocols.

To reduce bitrate further, \cite{diffusion_video_compression} employs diffusion models to choose which frames to transmit and to reconstruct inter-frame content. A conditional decoder generates subsequent frames from several compressed ones. While effective for compression, this design assumes error-free channels—unrealistic in wireless settings. It also increases decoder-side computation and, to meet a target quality, forces the encoder to emulate the decoder’s generative process when deciding whether to transmit or synthesize each frame, raising encoder-side complexity, latency, and energy consumption. Finally, the computational cost of diffusion sampling limits the reported resolution to $128\times 128$.

Recent advances in generative foundation models, such as Stable Diffusion \cite{Rombach_2022_CVPR} and Open-Sora \cite{opensora}, show promise for ultra-low-rate video transmission. By transmitting a text prompt, a first-frame reference, and minimal per-frame sketch cues, generative communication methods \cite{yin2025generativevideosemanticcommunication} can significantly reduce bitrate while maintaining semantic consistency. However, producing high-quality prompts typically requires a large video semantic extractor (e.g., the 7-billion-parameter Video-LLaVA model \cite{linetal2024videollava}). In addition, the computational cost of foundation models on the decoder side incurs substantial reconstruction latency, limiting their suitability for energy-constrained devices and latency-sensitive applications. Finally, separating sketch cues from the first-frame reference introduces redundancy that reduces compression efficiency.

These limitations motivate low-complexity and resilient solutions for ultra-low-bitrate video transmission that align with commodity digital hardware in energy-efficient deployment and sustain video recovery quality under adverse channel conditions.

Consequently, in this paper, we propose Resi\mbox{-}VidTok, a novel resilient tokenization-enabled framework for ultra-low-rate and lightweight video transmission by composing several cooperation-ready, low-complexity modules. Specifically, Resi\mbox{-}VidTok decomposes the complex video transmission system into three cooperating modules: (i) unified token-domain spatial--temporal compression with prefix-decodable reconstruction, (ii) frame sparsification with lightweight decoder-side interpolation, and (iii) channel-adaptive joint source--channel coding and modulation. By operating in a discrete and progressive token space and aligning transmission allocation with token importance and channel conditions, Resi\mbox{-}VidTok enables resilient delivery, graceful quality degradation, and coherent reconstruction using a single shared framewise decoder and real-time video interpolator, without an extra temporal neural extractor or generative model.

The main contributions are summarized as follows:
\begin{enumerate}
\item \textbf{Modular, resilience-aware architecture.} We decompose a monolithic video transmission system into lightweight modules centered on hybrid tokenization and temporal differencing, simplifying rate control, reducing computation, and enabling plug-and-play protection and decoding for robustness. Resi\mbox{-}VidTok performs well in visual and semantic consistency even at channel bandwidth ratios (CBR) as low as $4\times 10^{-4}$ and achieves video reconstruction exceeding $30$~fps.
\item \textbf{Unified token-domain compression across space--time.} We adapt a lightweight, framewise 1D tokenizer to video and integrate differential temporal token compression with zero-flag padding. The resulting importance-ordered bitstream supports prefix-decodable reconstruction from any token prefix via a single shared decoder, without auxiliary temporal models.
\item \textbf{Frame selection with lightweight interpolation.} Stride-controlled key-frame selection reduces transmission load, while a lightweight decoder-side interpolator reconstructs skipped frames, preserving motion continuity and visual fidelity at ultra-low rates.
\item \textbf{Channel-adaptive resilient delivery.} An importance-aware source coding strategy is proposed and coupled with channel-adaptive coding/modulation to allocate rate and modulation order by token importance and channel conditions, ensuring robust performance with graceful degradation across challenging SNRs.
\end{enumerate}

\section{System Model}

Consider wireless transmission of a video sequence
$\mathcal{X} := (\mathbf{x}_t)_{t=1}^{T}$, where each frame
$\mathbf{x}_t \in \mathbb{R}^{m}$ denotes a vectorized image (e.g., RGB
pixels) at time step $t$. We assume causal, low-latency operation: frames are encoded,
transmitted, and reconstructed sequentially as they arrive. Following a
GOP-like scheduling strategy \cite{deepjscc_v_2022tung,deepjscc_v_2023wang},
we partition $\mathcal{X}$ into contiguous groups of at most $N$ frames.
Let $G=\lceil T/N\rceil$ and define the $g$-th GOP as
\[
\mathcal{X}^{(g)} := \big\{\mathbf{x}_{(g-1)N+i}\big\}_{i=1}^{N_g},
\quad g=1,\ldots,G,
\]
where $N_g=\min\{N,\,T-(g-1)N\}$. When $T$ is a multiple of $N$, we have
$N_g=N$ for all $g$ and $T=GN$. This formulation captures the practical
GOP structure used for low-latency streaming while preserving strict
causality in both encoding and decoding.

The transmitter encodes the sequence into a stream of continuous-valued channel input vectors
\(\mathcal{S}=\{\mathbf{s}_t\}_{t=1}^{T}\), where \(\mathbf{s}_t\in\mathbb{R}^{k_t}\)
denotes the length-\(k_t\) channel input vector for video frame at time step \(t\). Typically
\(k_t<m\), and the average channel bandwidth ratio (CBR) \cite{deepjscc_v_2022tung} is utilized to evaluate the cost of communication bandwidth, which is defined as:
\begin{equation}
R \triangleq \frac{1}{T}\sum_{t=1}^{T} \frac{k_t}{m},
\label{eq:cbr}
\end{equation}
measuring the average number of channel symbols per source pixel.

The wireless channel is modeled by a (possibly stochastic) transfer function
\(\hat{\mathbf{s}}_t = W(\mathbf{s}_t;\boldsymbol{\nu})\) with parameters \(\boldsymbol{\nu}\) and an associated conditional law \(p_{\hat{\mathbf{s}}_t\,|\,\mathbf{s}_t}\). In this work, unless otherwise specified, we focus on the additive white
Gaussian noise (AWGN) channel,
\begin{equation}
\hat{\mathbf{s}}_t = \mathbf{s}_t + \mathbf{n}_t, \quad \mathbf{n}_t \sim \mathcal{N}\!\left(\mathbf{0},\, \sigma^2 \mathbf{I}_{k_t}\right),
\end{equation}
where \(\sigma^2\) denotes the (per-symbol) noise power and components of
\(\mathbf{n}_t\) are independent. Other channel models can be incorporated by
modifying \(W(\cdot;\boldsymbol{\nu})\).

The receiver applies the inverse processing chain to recover
\(\hat{\mathbf{x}}_t\) from \(\hat{\mathbf{s}}_t\) or to support downstream
intelligent tasks.

\section{Resi-VidTok Framework}

\subsection{Framework of Resi\mbox{-}VidTok}
\label{subsec:framework}
\begin{figure*}[t]
    \centering
    \includegraphics[width=0.98\textwidth]{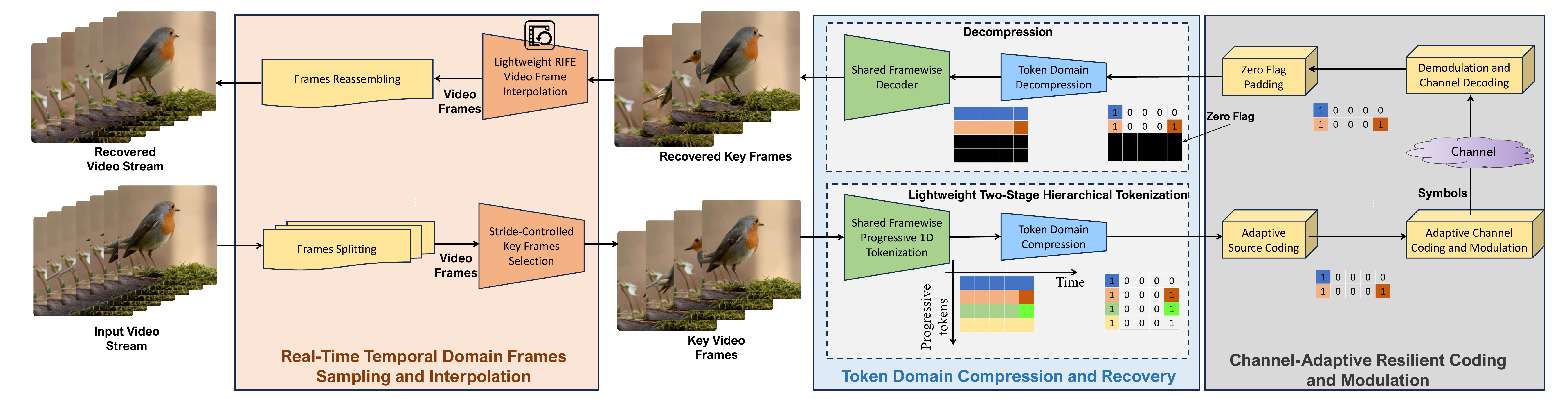}
    \caption{System overview of Resi-VidTok framework for ultra-low-rate robust video transmission.}
    \label{fig:system_overview}
\end{figure*}

We propose \emph{Resi\mbox{-}VidTok} as a lightweight, resilience-aware framework for end-to-end wireless video transmission that operates entirely in a discrete, progressive token space. As shown in Fig.~\ref{fig:system_overview}, the architecture has three cooperating subsystems:

\begin{itemize}
  \item \textbf{Real-time temporal-domain frame sampling \& recovery.}
  A stride-controlled selector partitions the input stream $\{\mathbf{x}_t\}_{t=1}^{T}$ into key frames $\mathcal{K}$ and non-key frames. Only key frames are transmitted; non-key frames are reconstructed at the receiver by a lightweight video-frame interpolator using neighboring recovered key frames, reducing bandwidth and reconstruction complexity.
  \item \textbf{Token-domain compression \& recovery.}
  Each selected key frame is mapped by a shared framewise progressive 1D tokenizer to a sequence of importance-ordered tokens; temporal correlation across \emph{consecutive key frames} is exploited via \emph{binary} differential masks, followed by rate-adaptive truncation and a compact header–body packing.
  \item \textbf{Channel-adaptive resilient coding \& modulation.}
  Based on the channel condition, a adaptive modulation–coding setting (MCS) is utilized to protect the transmission data and further coupled with source coding to determine the deliverable bit budget. At the receiver, demodulation/decoding is followed by zero-flag padding and prefix-decodable reconstruction.
\end{itemize}

Let $S$ be the key–frame stride and
$\mathcal{K}=\{1,1+S,1+2S,\ldots\}\cap\{1,\ldots,T\}$ the set of key indices.
When $t\in\mathcal{K}$ (a key frame), the end-to-end processing chain of
Resi\mbox{-}VidTok is summarized in \eqref{eq:rv_primary}. The current frame $\mathbf{x}_t$ is first mapped by the shared
framewise tokenizer $f_{\mathrm{tok}}$ to a sequence of importance–ordered
tokens $\mathbf{z}_t=(z_{t,1},\ldots,z_{t,L_t})$. To exploit temporal
redundancy across consecutive key frames, the \emph{differential token
compression} module $\Phi_{\mathrm{diff}}(\cdot\,|\,\mathbf{z}_{t_-})$ compares
$\mathbf{z}_t$ with the previous key–frame tokens
$\mathbf{z}_{t_-}$, where $t_-=\max\{k\in\mathcal{K}:k<t\}$, and constructs a
\emph{binary change mask} $\mathbf{m}_t$. Based on the instantaneous channel state, the adapter selects a MCS $(\rho_t,M_t)$ and the resulting deliverable bit budget $B_t$ (Sec.~\ref{subsec:channel_adapt_topk}); we then choose a prefix depth $K$ and restrict to the top-$K$ positions, forming a $K$-bit transmit mask $\mathbf{r}_t^{(K)}$ and an ordered list of updated token values $\mathbf{v}_t^{(K)}$ (values only where $\mathbf{r}_t^{(K)}$ has ones). The source packer $\mathcal{P}_{\mathrm{src}}$ then forms a compact bitstream $\mathbf{b}_t^{(K)}=\mathbf{h}_t^{(K)}\Vert\mathbf{v}_t^{(K)}$, where $\mathbf{h}_t^{(K)}=\mathrm{Enc}_{\mathrm{hdr}}(\mathbf{r}_t^{(K)})$ (the \emph{FrameHeader}) encodes which of the top-$K$ indices are updated and $\mathbf{v}_t^{(K)}=\mathrm{Enc}_{\mathrm{val}}(\cdot)$ (the \emph{FrameBody}) concatenates only the values of those updated tokens. The modulator/coder
$g_{\mathrm{m}}(\cdot;\rho_t,M_t)$ maps $\mathbf{b}_t^{(K)}$ to channel inputs
$\mathbf{s}_t$, which traverse the channel $W(\cdot;\bm{\nu})$ and are
demodulated/decoded as $\hat{\mathbf{b}}_t^{(K)}$ by
$g_{\mathrm{dem}}(\cdot;\rho_t,M_t)$. 

The unpacker $\mathcal{U}_{\mathrm{src}}$
parses $\hat{\mathbf{b}}_t^{(K)}$ into $(\hat{\mathbf{r}}_t^{(K)},\hat{\mathbf{v}}_t^{(K)})$ and
updates a running token state $\tilde{\mathbf{z}}_t$ by copying unchanged
positions from the previous state and overwriting only the indices indicated by
$\hat{\mathbf{r}}_t^{(K)}$; any missing entries (due to truncation or residual
errors) are filled by \emph{zero–flag padding} with a reserved zero-flag token.
Finally, the shared decoder $f_{\mathrm{dec}}$ reconstructs the frame
$\hat{\mathbf{x}}_t=f_{\mathrm{dec}}(\tilde{\mathbf{z}}_t)$. Because
$f_{\mathrm{dec}}$ is prefix–decodable and values in $\mathbf{v}_t^{(K)}$ are
ordered by importance, reconstruction quality improves monotonically as
more reliable bits are obtained.
\begin{equation}
    \label{eq:rv_primary}
    \begin{aligned}
    &\mathbf{x}_t
    \xrightarrow{\,f_{\mathrm{tok}}\,}
    \mathbf{z}_t
    \xrightarrow{\,\Phi_{\mathrm{diff}}(\cdot\,|\,\mathbf{z}_{t_-})\,}
    \big(\mathbf{r}_t^{(K)},\mathbf{v}_t^{(K)}\big)
    \xrightarrow{\,\mathcal{P}_{\mathrm{src}}\,}
    \mathbf{b}_t^{(K)} \\[2pt]
    \xrightarrow{\,g_{\mathrm{m}}(\cdot;\rho_t,M_t)\,}
    &\mathbf{s}_t
    \xrightarrow{\,W(\cdot;\bm{\nu})\,}
    \hat{\mathbf{s}}_t
    \xrightarrow{\,g_{\mathrm{dem}}(\cdot;\rho_t,M_t)\,}
    \hat{\mathbf{b}}_t^{(K)}
    \xrightarrow{\,\mathcal{U}_{\mathrm{src}}\,}
    \tilde{\mathbf{z}}_t
    \xrightarrow{\,f_{\mathrm{dec}}\,}
    \hat{\mathbf{x}}_t .
    \end{aligned}
\end{equation}

When $t\notin\mathcal{K}$ (a non–key frame), Resi\mbox{-}VidTok does not
transmit new tokens. Instead, the receiver synthesizes the frame by
interpolating between the nearest recovered key frames at indices
$t_-=\max\{k\in\mathcal{K}:k\le t\}$ and $t_+=\min\{k\in\mathcal{K}:k\ge t\}$:
\begin{equation}
\label{eq:rv_interp}
\hat{\mathbf{x}}_t \;=\; f_{\mathrm{int}}\!\left(
\hat{\mathbf{x}}_{t_-},\,\hat{\mathbf{x}}_{t_+},\,\alpha_t\right),
\qquad \alpha_t=\frac{t-t_-}{t_+-t_-},
\end{equation}
with standard boundary handling at the ends of each GOP. 

\subsection{Real\mbox{-}time Temporal Domain Frame Sampling}
\label{subsec:rt_sampling}

To reduce encoder workload and early remove temporal redundancy, Resi\mbox{-}VidTok performs causal frame sampling before tokenization. Let $S\!\in\!\mathbb{N}$ be the key\mbox{-}frame stride and define the key\mbox{-}frame index set
\[
\mathcal{K} \;=\; \{1,\,1+S,\,1+2S,\ldots\}\cap\{1,\ldots,T\}.
\]
Only frames with indices in $\mathcal{K}$ are passed to $f_{\mathrm{tok}}$ and subsequently transmitted; frames with $t\notin\mathcal{K}$ are reconstructed at the receiver (Sec.~\ref{subsec:rt_recovery}). When $S\!=\!1$, the sampler becomes the identity and every frame is encoded and transmitted, recovering the no\mbox{-}sampling baseline.

This stride\mbox{-}controlled selector is attractive for two reasons. \emph{(i) Computational savings:} the number of tokenizer invocations scales with $|\mathcal{K}|\!=\!\lceil T/S\rceil$, so the encoder complexity drops from $T$ forward passes to $\lceil T/S\rceil$ without modifying the tokenizer. \emph{(ii) Early temporal compression:} by transmitting only $1/S$ of the frames (in the uniform case), we remove a large portion of temporal redundancy before token formation; the remaining redundancy between \emph{consecutive key frames} is then handled by differential token compression in Sec.~\ref{subsubsec:dtc}. 

The stride $S$ can be chosen based on application latency and channel budget. In this paper we use a fixed $S$ for clarity and fair comparison; however, the module is \emph{policy\mbox{-}agnostic}. Future work can replace the uniform selector by a more advanced policy (e.g., content/motion\mbox{-}aware, channel\mbox{-}aware, or learning\mbox{-}to\mbox{-}sample) that adapts $S$ or selects non\mbox{-}uniform key indices to further improve rate\mbox{--}distortion\mbox{--}latency trade\mbox{-}offs.

\subsection{Lightweight Two\mbox{-}Stage Token\mbox{-}Domain Compression}
\label{subsec:two_stage_token}

\begin{figure}[t]
    \centering
    \includegraphics[width=\columnwidth]{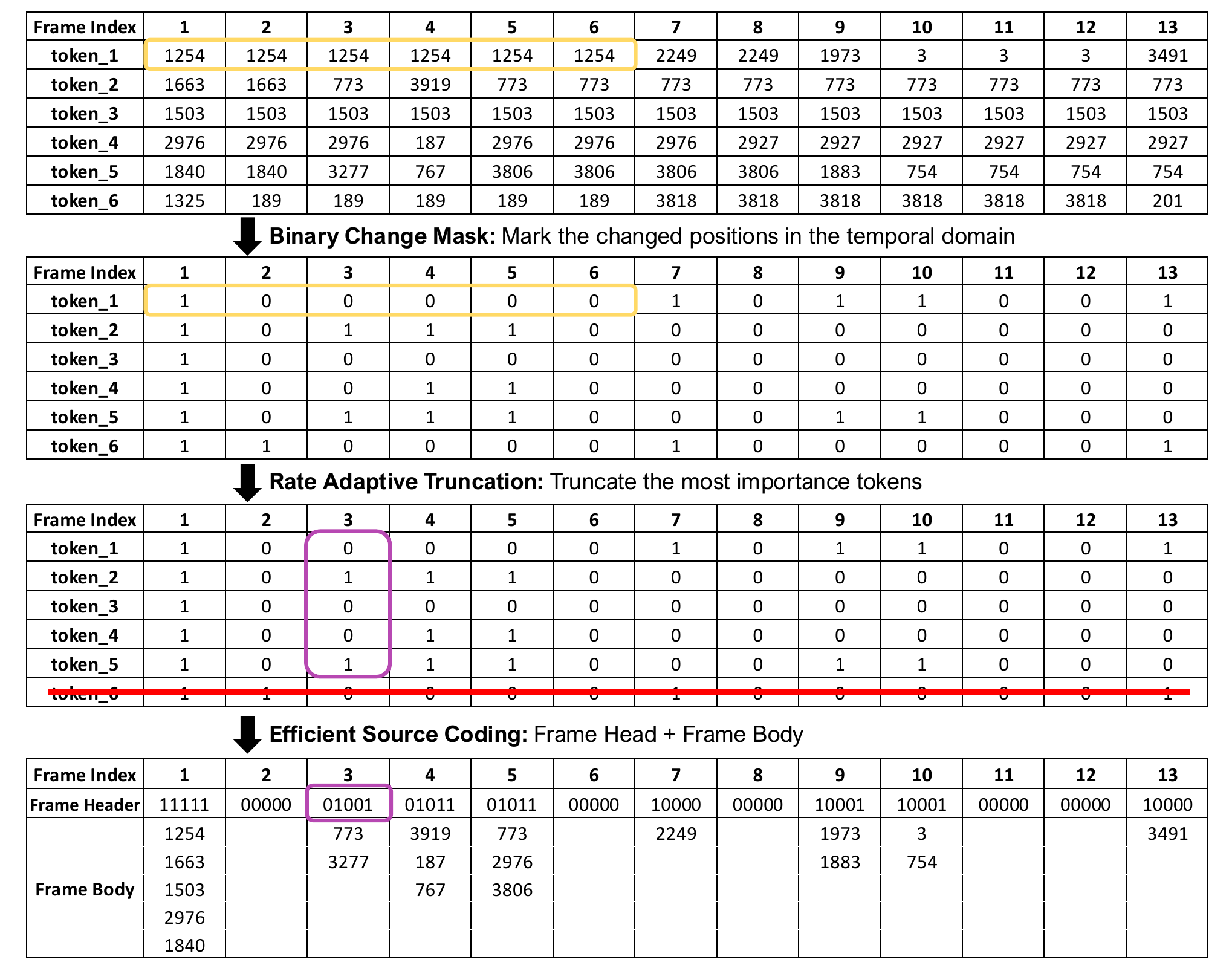}
    \caption{Example of two-stage token-domain compression and adaptive source coding.}
    \vspace{-10pt}
    \label{figure_transnet}
\end{figure}

Resi\mbox{-}VidTok performs compression entirely in a \emph{discrete} token space with two lightweight stages: (i) a \emph{shared} framewise tokenizer that yields importance\mbox{-}ordered, prefix\mbox{-}decodable tokens for every key frame, and (ii) a \emph{binary} differential mechanism that converts temporal redundancy between consecutive key frames into sparse updates. This design deliberately avoids heavy temporal modules (e.g., attention across long clips \cite{deepjscc_v_2022tung,yin2025generativevideosemanticcommunication}, or generative decoders \cite{diffusion_video_compression,yin2025generativevideosemanticcommunication}), thereby reducing complexity and latency while preserving compatibility with the image case and enabling progressive refinement.

\subsubsection{Efficient Progressive Shared Framewise Tokenizer}
\label{subsubsec:shared_tokenizer}

Given a key video frame $\mathbf{x}_t\!\in\!\mathbb{R}^{H\times W\times C}$, a single lightweight tokenizer $f_{\mathrm{tok}}$ (reused for all frames) produces a token string
\begin{equation}
\mathbf{z}_t \;=\; f_{\mathrm{tok}}(\mathbf{x}_t)\;=\;(z_{t,1},\ldots,z_{t,L_t}), 
\qquad z_{t,\ell}\in\mathcal{V},
\label{eq:tokenizer_lw2}
\end{equation}
where $\mathcal{V}$ is a finite vocabulary and $L_t$ is the per\mbox{-}frame token length (padded to a GOP\mbox{-}wise maximum if needed). Tokens are ordered by importance scores $w_{t,\ell}\!\in\!\mathbb{R}_+$ so that $w_{t,1}\!\ge\!\cdots\!\ge\!w_{t,L_t}$. A shared decoder $f_{\mathrm{dec}}$ is \emph{prefix\mbox{-}decodable}, i.e.,
\begin{equation}
\hat{\mathbf{x}}_t^{(\ell)} \;=\; f_{\mathrm{dec}}\!\big(z_{t,1:\ell}\big), 
\qquad \ell=0,1,\ldots,L_t,
\label{eq:prefix_decode_lw2}
\end{equation}
and the reconstruction quality improves monotonically with $\ell$. Following zero\mbox{-}out training as in ResiTok \cite{liu2025resitokresilienttokenizationenabledframework}, early tokens capture semantics (“key” tokens) while later tokens refine details, enabling graceful degradation under erasures or truncation. Notably, we do not introduce an explicit video tokenizer, and temporal adaptation is delegated to a binary differential mechanism downstream, keeping the tokenizer image\mbox{-}compatible and computationally lean.

\subsubsection{Differential Token Compression (Binary Mask Across Key Frames)}
\label{subsubsec:dtc}

Let $\mathcal{K}$ be the key\mbox{-}frame index set (stride $S$). For a current key frame $t\!\in\!\mathcal{K}$ and the previous key index $t_-=\max\{k\in\mathcal{K}:k<t\}$, we detect token\mbox{-}level changes relative to $\mathbf{z}_{t_-}$ using a \emph{binary} change mask:
\begin{equation}
m_{t,\ell} \;=\; \mathbbm{1}\!\left[z_{t,\ell}\neq z_{t_-,\ell}\right], 
\qquad \mathbf{m}_t\in\{0,1\}^{L_t}.
\label{eq:mask_lw2}
\end{equation}
The candidate update set is $\mathcal{C}_t=\{\ell: m_{t,\ell}=1\}$. Because mask construction is a single pass over tokens, complexity is $O(L_t)$ with negligible memory. By reducing temporal modeling to a deterministic binary mask, we convert motion/appearance changes into \emph{sparsity} at the token level, which (i) eliminates learned temporal motion modules, (ii) enables progressive updates driven by importance, and (iii) prepares a \emph{structure\mbox{-}aware} payload for the channel interface. The mask and associated changed values feed the rate controller in Sec.~\ref{subsec:channel_adapt_topk}.

\subsection{Channel\mbox{-}adaptive Resilient Coding \& Modulation}
\label{subsec:channel_adapt_topk}

We couple PHY adaptation with token\mbox{-}domain source decisions by first estimating how many bits can be reliably delivered under the current channel and the corresponding MCS, and then choosing how many \emph{top\mbox{-}K tokens} to materialize. To save the coding bits, here we \emph{always} consider the first $K$ tokens and use a \emph{binary header of length $K$} to indicate which of those $K$ tokens have changed. Consequently, the header cost is exactly $K$ bits, while the body cost equals a constant value length per token times the number of ones in the header.

\paragraph{MCS selection and per\mbox{-}frame deliverable bits.}
For each transmitted key frame $t\in\mathcal{K}$, the adapter estimates channel quality $\hat{\gamma}_t$ and selects a single MCS from a standard table to meet a target BLER $\varepsilon$:
\begin{equation}
(\rho_t,M_t) \;=\; \mu(\hat{\gamma}_t) \quad
\text{s.t.}\;\; \mathrm{BLER}\big(\rho_t,M_t;\hat{\gamma}_t,k_t\big)\le\varepsilon .
\label{eq:mcs_select_topk}
\end{equation}
Given the planned channel uses $k_t$ (chosen under the global CBR $R$, cf. Eq.~\eqref{eq:cbr}), the \emph{deliverable} source bits for frame $t$ are
\begin{equation}
B_t \;=\; \Big\lfloor \rho_t\,k_t\,\log_2 M_t \Big\rfloor .
\label{eq:deliverable_bits_topk}
\end{equation}

\paragraph{Top\mbox{-}K tokens with binary header.}
Let $\mathbf{z}_t=(z_{t,1},\ldots,z_{t,L_t})$ be the importance\mbox{-}ordered tokens for frame $t$, and let $\mathbf{m}_t\!\in\!\{0,1\}^{L_t}$ be the binary change mask relative to the reference. For a chosen $K\!\in\!\{0,\ldots,L_t\}$, we form a \emph{$K$-limited transmit header} over the top-$K$ positions,
\begin{equation}
\mathbf{r}_t^{(K)} \;=\; \big(m_{t,1},\ldots,m_{t,K}\big)\in\{0,1\}^{K},
\label{eq:k_limited_mask}
\end{equation}
and encode it verbatim as the \emph{FrameHeader}:
\begin{equation}
\mathbf{h}_t^{(K)} \;=\; \mathrm{Enc}_{\mathrm{hdr}}\!\big(\mathbf{r}_t^{(K)}\big), 
\qquad \big|\mathbf{h}_t^{(K)}\big| \;=\; K \;\;\text{bits}.
\label{eq:head_len_is_K}
\end{equation}
Only those of the $K$ tokens that changed (i.e., $r_{t,i}^{(K)}\!=\!1$) contribute values to the \emph{FrameBody}. With a fixed per\mbox{-}token value length $b_{\mathrm{val}}$ (e.g., $b_{\mathrm{val}}\!=\!\lceil\log_2|\mathcal{V}|\rceil$ or a fixed\mbox{-}point code), the body length is
\begin{equation}
\big|\mathbf{v}_t^{(K)}\big| \;=\; b_{\mathrm{val}}\,
C_t(K), \quad C_t(K)\;\triangleq\;\sum_{i=1}^{K} m_{t,i},
\label{eq:body_len_topk}
\end{equation}
and the total source bits for frame $t$ under top\mbox{-}$K$ are
\begin{equation}
n_t(K) \;=\; \underbrace{K}_{\text{header}} \;+\; 
\underbrace{b_{\mathrm{val}}\,C_t(K)}_{\text{body}} .
\label{eq:frame_cost_topk}
\end{equation}
\emph{Convention for the first transmitted frame in a GOP:} to bootstrap the token state when no within\mbox{-}GOP reference exists, set $m_{t,i}\!\equiv\!1$ (or compare to a zero state), so $C_t(K)\!=\!K$ and $n_t(K)\!=\!(1\!+\!b_{\mathrm{val}})K$.

\paragraph{Budget\mbox{-}feasible top\mbox{-}$K$ per frame (fast search).}
Given $B_t$ from \eqref{eq:deliverable_bits_topk} and the mask prefix sums
$C_t(K)$, the \emph{largest} feasible token count for frame $t$ is
\begin{equation}
K_t^\star \;=\; \max\Big\{K\in\{0,\ldots,L_t\}\,:\; n_t(K)\le B_t\Big\}.
\label{eq:Kt_star_topk}
\end{equation}
Because $n_t(K)$ is monotone nondecreasing in $K$ (each increment adds one
header bit and possibly $b_{\mathrm{val}}$ if $m_{t,K}\!=\!1$), $K_t^\star$ can be
found via a binary search after an $O(L_t)$ precomputation of $C_t(K)$.

\paragraph{Optional GOP\mbox{-}level planning (stabilized prefix).}
If consistent prefix depth within a GOP is desired, select a single
$K^{(g)}$ for all transmitted key frames of GOP $g$:
\begin{equation}
    K^{(g)} \;=\; \min_{t\in\mathcal{T}^{(g)}} K_t^\star ,
    \label{eq:Kg_topk}
    \end{equation}
ensuring $n_t\!\big(K^{(g)}\big)\!\le\!B_t$ for every transmitted key frame while
keeping a stable prefix length, where $\mathcal{K}^{(g)} \;\triangleq\; \mathcal{K} \cap \{(g{-}1)N+1,\ldots,(g{-}1)N+N_g\}$. The corresponding GOP cost is
$\sum_{t\in\mathcal{K}^{(g)}} n_t\!\big(K^{(g)}\big)$, fully determined by
$K^{(g)}$ and $C_t\!\big(K^{(g)}\big)$.

\paragraph{PHY mapping and decoding.}
Once $K$ (either $K_t^\star$ or $K^{(g)}$) is fixed, we pack
$\mathbf{b}_t^{(K)}=\mathbf{h}_t^{(K)}\Vert\mathbf{v}_t^{(K)}$ with
$|\mathbf{b}_t^{(K)}|=n_t(K)\le B_t$, map it with the selected MCS, and send:
\begin{subequations}
    \label{eq:phy_map_topk}
    \begin{align}
    \mathbf{s}_t \;=\; g_{\mathrm{m}}\!\big(\mathbf{b}_t^{(K)};\rho_t,M_t\big), \label{eq:phy_map_topk:a}\\
    \hat{\mathbf{s}}_t \;=\; W(\mathbf{s}_t;\boldsymbol{\nu}). \label{eq:phy_map_topk:b}
    \end{align}
\end{subequations}

\subsection{Reception and Reconstruction}
\label{subsec:reception}

This subsection details the receiver pipeline for transmitted (key or sampled) frames under the top\mbox{-}$K$ token scheme of Sec.~\ref{subsec:channel_adapt_topk}, and completes the loop from demodulation to detokenization. The goal is to reliably parse the header/body produced by the source coder, update a running token state using only the tokens indicated by the header, and reconstruct the video frame via a shared, prefix\mbox{-}decodable decoder.

\paragraph{Demodulation and channel decoding.}
For a transmitted frame $t$, the baseband vector $\hat{\mathbf{s}}_t$ is demodulated and decoded using the selected MCS $(\rho_t,M_t)$ (Sec.~\ref{subsec:channel_adapt_topk}) to produce a bitstream estimate
\begin{equation}
\hat{\mathbf{b}}_t^{(K)} \;=\; g_{\mathrm{dem}}\!\big(\hat{\mathbf{s}}_t;\rho_t,M_t\big)
\;=\; \hat{\mathbf{h}}_t^{(K)} \,\Vert\, \hat{\mathbf{v}}_t^{(K)},
\label{eq:rx_bitstream}
\end{equation}
where  $\hat{\mathbf{h}}_t^{(K)}$ and  $\hat{\mathbf{v}}_t^{(K)}$ are recovered header and body. A light CRC (or parity) over $\hat{\mathbf{b}}_t^{(K)}$ can be used to detect residual frame\mbox{-}level errors. Upon failure, Resi\mbox{-}VidTok falls back to a \emph{no\mbox{-}update} policy for frame $t$ (the running token state remains unchanged), preserving temporal coherence and avoiding HARQ/ARQ latency.

\paragraph{Header parsing and value extraction.}
The header is decoded verbatim to recover the $K$\mbox{-}bit transmit mask over the top\mbox{-}$K$ positions:
\begin{equation}
\hat{\mathbf{r}}_t^{(K)} \;=\; \mathrm{Dec}_{\mathrm{hdr}}\!\big(\hat{\mathbf{h}}_t^{(K)}\big)
\;\in\; \{0,1\}^{K},
\qquad
\hat{r}_{t,i}^{(K)} \in \{0,1\}.
\label{eq:hdr_parse_rx}
\end{equation}
Let $\mathrm{ReadVal}(\cdot)$ be a pointer that consumes the next $b_{\mathrm{val}}$ bits from $\hat{\mathbf{v}}_t^{(K)}$ and maps them back to a token value in $\mathcal{V}$. The number of values present equals the number of ones in the header, $C_t(K)\!=\!\sum_{i=1}^{K}\hat{r}_{t,i}^{(K)}$.

\paragraph{Running token state update.}
Let $\tilde{\mathbf{z}}_{t-1}=(\tilde{z}_{t-1,1},\ldots,\tilde{z}_{t-1,L_t})$ denote the receiver’s token state \emph{just before} processing frame $t$. For the first transmitted frame of a GOP, initialize $\tilde{\mathbf{z}}_{t-1}$ to a zero state $\mathbf{z}_{\mathrm{zero}}$ with all entries equal to a reserved token $z_{\mathrm{zero}}\!\in\!\mathcal{V}$. Then update
\begin{equation}
    \label{eq:token_state_rx}
    \tilde{z}_{t,i} \;=\;
    \begin{cases}
    \mathrm{ReadVal}\!\big(\hat{\mathbf{v}}_t^{(K)}\big), &
    \begin{aligned}[t]
    &1\le i\le K,\\[-1pt]
    &\hat{r}_{t,i}^{(K)}=1,
    \end{aligned}\\[4pt]
    \tilde{z}_{t-1,i}, &
    \begin{aligned}[t]
    &1\le i\le K,\\[-1pt]
    &\hat{r}_{t,i}^{(K)}=0,
    \end{aligned}\\[4pt]
    \tilde{z}_{t-1,i}, &
    \begin{aligned}[t]
    &i>K,
    \end{aligned}
    \end{cases}
    \qquad i=1,\ldots,L_t.
    \end{equation}
    
If the header indicates an update position but the body is shorter than expected (rare under CRC), replace the missing value by $z_{\mathrm{zero}}$ (zero\mbox{-}flag padding) to keep the sequence well\mbox{-}formed.

\paragraph{Detokenization and progressive reconstruction.}
The updated state $\tilde{\mathbf{z}}_t$ feeds the shared, prefix\mbox{-}decodable decoder $f_{\mathrm{dec}}$ (Sec.~\ref{subsubsec:shared_tokenizer}):
\begin{equation}
\hat{\mathbf{x}}_t \;=\; f_{\mathrm{dec}}\!\big(\tilde{\mathbf{z}}_t\big).
\label{eq:detok_recon}
\end{equation}
Because tokens are ordered by importance and only the top\mbox{-}$K$ positions are eligible for change per transmitted frame, \eqref{eq:token_state_rx} effectively realizes a \emph{stable prefix} whose depth is controlled by $K$ (either $K_t^\star$ or a GOP\mbox{-}wide $K^{(g)}$). As the channel permits larger $B_t$ (Sec.~\ref{subsec:channel_adapt_topk}), the chosen $K$ increases, more high\mbox{-}impact tokens are materialized, and reconstruction quality improves monotonically. For unchanged positions and indices beyond $K$, inheritance from $\tilde{\mathbf{z}}_{t-1}$ maintains temporal consistency of semantics and texture without re\mbox{-}sending redundant tokens.

\paragraph{Complexity and latency.}
Receiver overhead is linear in $K$: parsing a $K$\mbox{-}bit header, reading up to $C_t(K)$ token values, and performing $O(L_t)$ state copies (implemented as in\mbox{-}place updates over a ring buffer). The decoder $f_{\mathrm{dec}}$ is shared with the image case and is invoked once per transmitted frame. Non\mbox{-}transmitted frames are recovered by interpolation in Sec.~\ref{subsec:rt_recovery}, so no detokenization is needed for those frames. Overall, the reception path retains the standard PHY stack (single MCS per frame) and achieves resilience via simple, deterministic parsing and prefix\mbox{-}based detokenization in the token domain.

\subsection{Real\mbox{-}time Temporal Domain Frame Recovery}
\label{subsec:rt_recovery}

Non\mbox{-}key frames are reconstructed at the receiver from neighboring decoded \emph{key frames}. Denote by
\[
t_-=\max\{k\in\mathcal{K}:k\le t\},\qquad 
t_+=\min\{k\in\mathcal{K}:k\ge t\}
\]
the nearest key indices surrounding a non\mbox{-}key time $t$, and let $\alpha_t=\frac{t-t_-}{t_+-t_-}\in[0,1]$ be the normalized time. We instantiate the interpolation operator $f_{\mathrm{int}}$ by the \emph{Real\mbox{-}Time Intermediate Flow Estimation} (RIFE) network \cite{huang2022rife}, a fast optical\mbox{-}flow\mbox{-}based method that estimates bidirectional flows and occlusion masks and fuses warped features in a single forward pass. The reconstruction is
\begin{equation}
\hat{\mathbf{x}}_t \;=\; f_{\mathrm{int}}\!\big(\hat{\mathbf{x}}_{t_-},\,\hat{\mathbf{x}}_{t_+},\,\alpha_t;\,\phi_{\text{RIFE}}\big),
\label{eq:rt_interp_rife}
\end{equation}
where $\phi_{\text{RIFE}}$ are fixed network parameters. In practice, $f_{\mathrm{int}}$ runs once per missing time index and introduces only a bounded delay (at most $S\!-\!1$ frames) to wait for $\hat{\mathbf{x}}_{t_+}$, preserving low latency.

This design has several advantages aligned with the overall goal of low complexity and robustness: \emph{(i) Division of labor:} temporal synthesis for non\mbox{-}key frames is delegated to a dedicated, highly optimized video interpolation model from the vision community, allowing the communication pipeline to remain lightweight and token\mbox{-}centric. \emph{(ii) Encoder savings:} because only key frames are tokenized and transmitted, we substantially reduce encoder compute, rate, and energy. \emph{(iii) Progressive quality:} as key frames are refined by more reliable token prefixes (Sec.~\ref{subsubsec:shared_tokenizer}), the quality of $\hat{\mathbf{x}}_{t_-}$ and $\hat{\mathbf{x}}_{t_+}$ improves, which directly benefits the interpolated non\mbox{-}key frames through \eqref{eq:rt_interp_rife}. \emph{(iv) Graceful degradation:} when the channel budget is tight, the system can increase $S$ (fewer key frames) while still delivering temporally coherent videos; conversely, when the channel improves, $S$ can be reduced toward the identity case $S\!=\!1$.

Boundary handling at the beginning and end of a GOP follows standard practice: if $t_-$ (or $t_+$) is unavailable, we use simple propagation from the available side (copy or extrapolation) or defer interpolation until the next key frame arrives, depending on latency constraints. Overall, combining stride\mbox{-}based sampling with RIFE\mbox{-}based recovery yields a practical, real\mbox{-}time path to reconstruct dense video from sparse transmitted key frames, and complements our token\mbox{-}domain design to keep the entire system low\mbox{-}complexity, energy\mbox{-}efficient, and resilient.

\section{Experimental Results}

\subsection{Experimental Setup}
\label{subsec:exp_setup}

\paragraph{Datasets and preprocessing.}
Following prior work \cite{diffusion_video_compression,deepjscc_v_2023wang}, we evaluate on the UVG dataset \cite{uvg_dataset}, which contains seven high\mbox{-}resolution videos with diverse motion patterns. All frames are center\mbox{-}cropped and down\mbox{-}sampled to $256\times256$. Different from \cite{diffusion_video_compression}, which evaluates only the first $30$ frames per sequence, we evaluate on \emph{entire} videos. We also include WebVid \cite{webvid_dataset} as used in \cite{yin2025generativevideosemanticcommunication}: we randomly select $50$ videos longer than $20$\,s, keep the first $20$\,s at $30$\,fps, and evaluate on the \emph{first $300$ frames} of each video. Compared with \cite{yin2025generativevideosemanticcommunication} which uses first 8 frames of each video for evaluation, we use 300 frames of each video for evaluation. Unless stated otherwise, the GOP size is $N=32$.

\paragraph{Tokenizer and interpolation modules.}
We adopt the progressive, resilience\mbox{-}aware image tokenizer from \cite{liu2025resitokresilienttokenizationenabledframework} (trained on ImageNet \cite{ILSVRC15}) as the \emph{shared} framewise tokenizer $f_{\mathrm{tok}}$ for all video frames. During training of the tokenizer in \cite{liu2025resitokresilienttokenizationenabledframework}, images are randomly cropped to $256\times256$ patches; we reuse the released weights without additional video\mbox{-}specific finetuning, thereby preserving the image/video alignment of our token space. Decoder\mbox{-}side interpolation of non\mbox{-}key frames is implemented with the lightweight Practical\mbox{-}RIFE \cite{huang2022rife} (Sec.~\ref{subsec:rt_recovery}), which provides real\mbox{-}time intermediate flow estimation and enables low\mbox{-}complexity temporal recovery on commodity hardware. The bits per token is set to 12.

\paragraph{Channel model and ACM configuration.}
Unless otherwise noted, we simulate an AWGN channel with noise variance chosen to realize the target SNR (in dB). Channel coding uses LDPC. Per transmitted frame, the receiver estimates SNR and the channel adapter selects a single MCS $(\rho_t,M_t)$ from a 3GPP\mbox{-}style table \cite{3GPP_38_214} such that the block error rate (BLER) does not exceed $0.002$. The corresponding code rate and modulation order then determine the deliverable bit budget  used by our top\mbox{-}$K$ token planner (Sec.~\ref{subsec:channel_adapt_topk}). The specific ACM levels used in our simulations are listed in Table~\ref{tab:acm}.

\begin{table}[htbp]
    \centering
    \caption{Adaptive Coding and Modulation (ACM) configurations used in simulations (AWGN channel, LDPC coding).}
    \begin{tabular}{c c c c}
    \toprule
    \textbf{SNR (dB)} & \textbf{Code Rate} & \textbf{Modulation} & \textbf{Order (bits/sym)} \\
    \midrule
    $-2$  & 0.245 & QPSK  & 2 \\
    $0$   & 0.301 & QPSK  & 2 \\
    $2$   & 0.514 & QPSK  & 2 \\
    $4$   & 0.663 & QPSK  & 2 \\
    $6$   & 0.424 & 16QAM & 4 \\
    $8$   & 0.540 & 16QAM & 4 \\
    $10$  & 0.643 & 16QAM & 4 \\
    \bottomrule
    \end{tabular}
    \label{tab:acm}
\end{table}

\paragraph{Baselines.}
Following \cite{deepjscc_v_2022tung,deepjscc_v_2023wang,yin2025generativevideosemanticcommunication}, we compare Resi\mbox{-}VidTok against a classical separated pipeline that uses HEVC/H.265 \cite{h265} for source coding and practical LDPC codes \cite{ldpc} for channel coding over the same AWGN channel and ACM table. For fairness, all methods are evaluated under matched resolution ($256\times256$) and the same GOP structure ($N=32$).


\subsection{Performance Comparison under Different CBR}
\label{subsec:cbr_comparison}

\begin{figure}[!t]
    \centering
    \subfigure[WebVid - CLIP Score]{
        \includegraphics[width=0.22\textwidth]{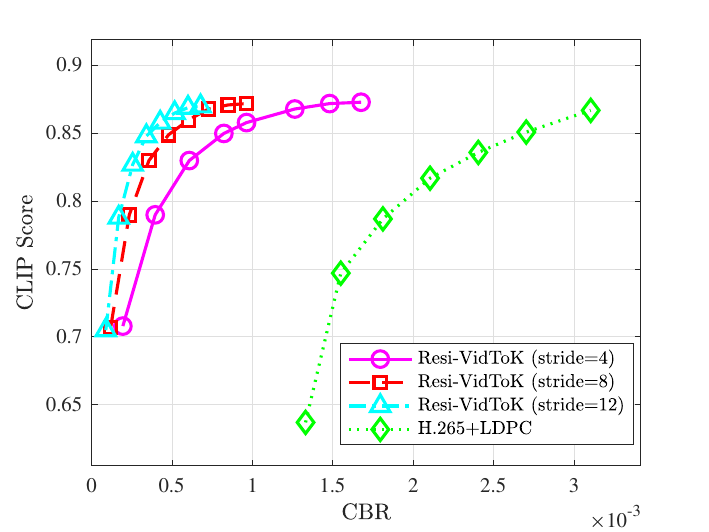}
    }
    \hfill
    \subfigure[UVG - CLIP Score]{
        \includegraphics[width=0.22\textwidth]{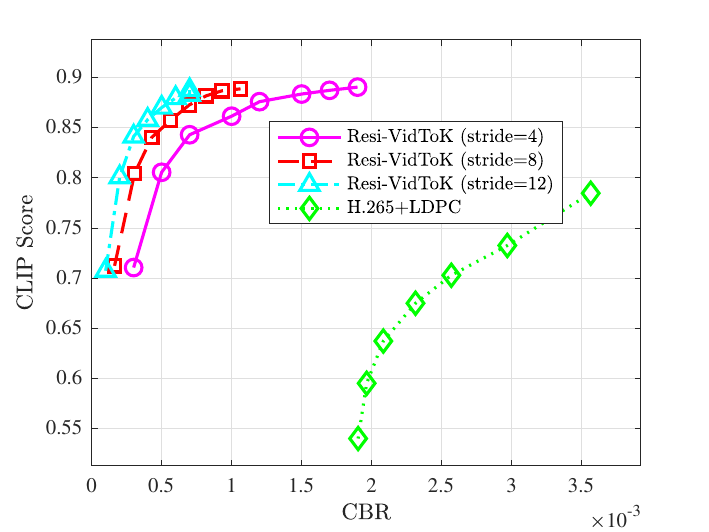}
    }
    \\ \vspace*{-3.9mm}
    
    \subfigure[WebVid - PSNR]{
        \includegraphics[width=0.22\textwidth]{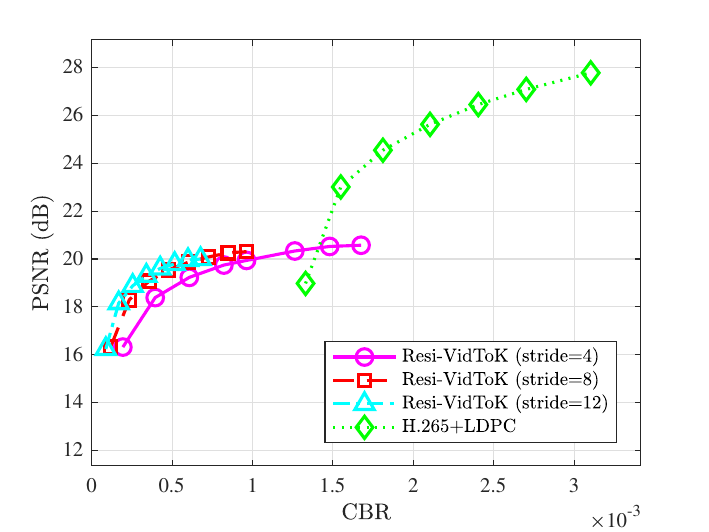}
    }
    \hfill
    \subfigure[UVG - PSNR]{
        \includegraphics[width=0.22\textwidth]{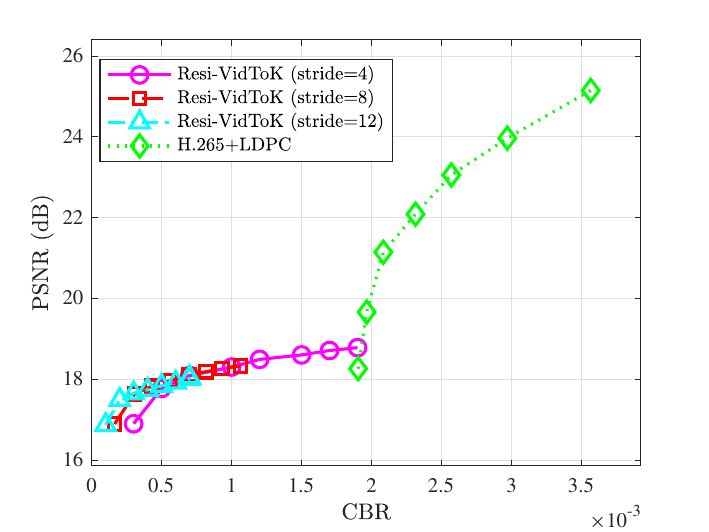}
    }
    \\ \vspace*{-3.9mm}
    
    \subfigure[WebVid - LPIPS]{
        \includegraphics[width=0.22\textwidth]{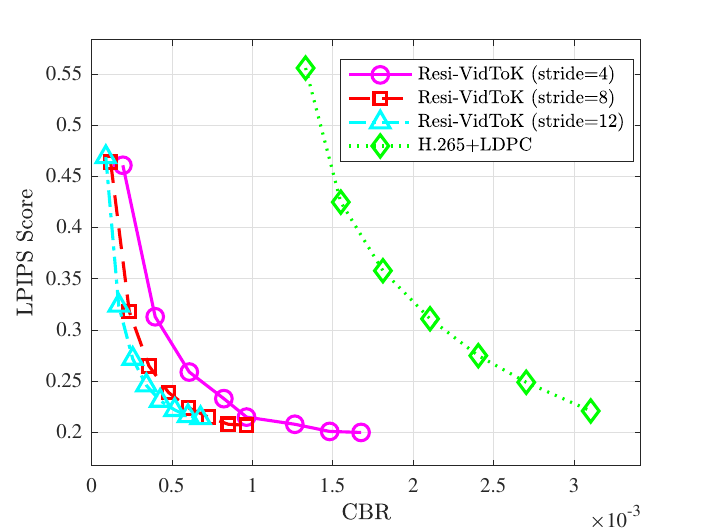}
    }
    \hfill
    \subfigure[UVG - LPIPS]{
        \includegraphics[width=0.22\textwidth]{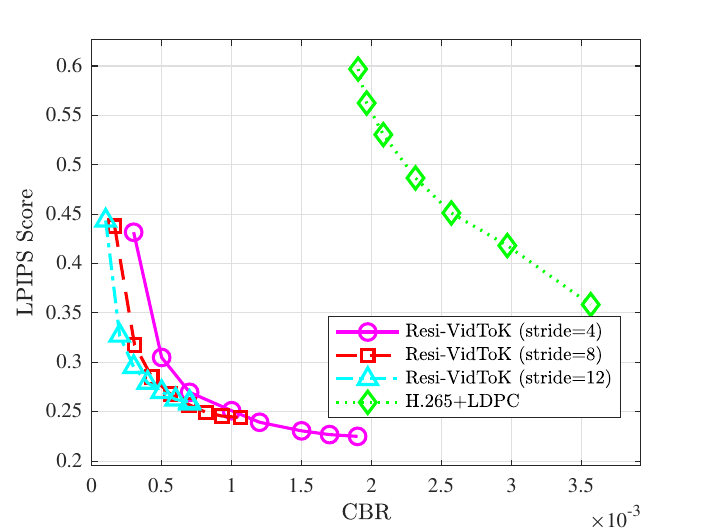}
    }
    
    \caption{Quality versus CBR at SNR=6 dB.}
    \vspace*{-3mm}
    \label{fig:cbr_comparison}
\end{figure}

Fig.~\ref{fig:cbr_comparison} reports quality versus CBR at $\mathrm{SNR}=6$\,dB with 16\mbox{-}QAM (BLER target $0.002$; ACM per Table~\ref{tab:acm}), comparing Resi\mbox{-}VidTok with three sampling strides $S\!\in\!\{4,8,12\}$ against the classical separated pipeline ``H.265\,+\,LDPC'' on WebVid and UVG. We evaluate semantic similarity (CLIP Score; higher is better), structural fidelity (PSNR; higher is better), and perceptual similarity (LPIPS; lower is better).

\paragraph{Semantic fidelity (CLIP).}
Across both datasets (Figs.~\ref{fig:cbr_comparison}a–b), Resi\mbox{-}VidTok achieves consistently higher CLIP scores than H.265\,+\,LDPC over the entire CBR range. Notably, Resi\mbox{-}VidTok reaches a high semantic plateau at very low rates (around $5\times10^{-4}$ and below), whereas the separated baseline needs substantially larger CBR to approach comparable levels. This reflects the importance\mbox{-}ordered tokenization and our channel\mbox{-}adaptive, top\mbox{-}$K$ delivery: the most informative tokens are delivered first and preserved even when the bit budget is tight, maintaining semantic integrity.

\paragraph{Perceptual quality (LPIPS).}
Resi\mbox{-}VidTok also dominates in LPIPS on both WebVid and UVG (Figs.~\ref{fig:cbr_comparison}e–f). The curves drop sharply and stabilize at lower LPIPS than H.265\,+\,LDPC across all tested CBRs, indicating better perceptual realism under extreme compression. This advantage stems from two factors: (i) stable prefix decoding of high\mbox{-}impact tokens yields coherent textures even when many refinements are omitted; and (ii) decoder\mbox{-}side interpolation reuses high\mbox{-}quality key\mbox{-}frame anchors to synthesize intermediate content.

\paragraph{Structure (PSNR).}
For PSNR (Figs.~\ref{fig:cbr_comparison}c–d), Resi\mbox{-}VidTok is competitive and often superior at ultra\mbox{-}low CBR (left side of the plots). As CBR increases, the MSE\mbox{-}oriented H.265\,+\,LDPC curves continue to grow and may surpass Resi\mbox{-}VidTok in absolute PSNR. This trade\mbox{-}off is expected: our design prioritizes semantic/perceptual preservation under tight budgets via token importance and GOP\mbox{-}level planning, while the baseline favors distortion reduction given sufficient bits. Importantly, even in regions where the baseline reaches higher PSNR, Resi\mbox{-}VidTok maintains superior CLIP and LPIPS, aligning better with human perception and downstream tasks.

\paragraph{Effect of sampling stride $S$.}
Varying the stride controls how often the tokenizer is invoked and how much channel budget is concentrated per transmitted frame. Smaller strides ($S{=}4$) generally yield slightly better LPIPS/PSNR at a fixed CBR, thanks to more frequent anchor refresh. Larger strides ($S{=}12$) can marginally help CLIP at the lowest CBRs by concentrating budget on fewer anchors. Overall, $S{=}8$ provides a balanced trade\mbox{-}off across all three metrics, consistent with our real\mbox{-}time, low\mbox{-}complexity objectives.

In summatry, under the same ACM and channel conditions, Resi\mbox{-}VidTok delivers \emph{higher semantic and perceptual quality} than the separated H.265\,+\,LDPC pipeline across a wide CBR range, and exhibits graceful degradation at ultra\mbox{-}low rates. These gains validate the core design choices—importance\mbox{-}ordered tokenization, channel\mbox{-}adaptive top\mbox{-}$K$ transmission, and lightweight interpolation—while keeping the PHY standard and the end\mbox{-}to\mbox{-}end latency low.

\begin{figure}[!t]
    \centering
    \subfigure[WebVid - CLIP Score]{
        \includegraphics[width=0.22\textwidth]{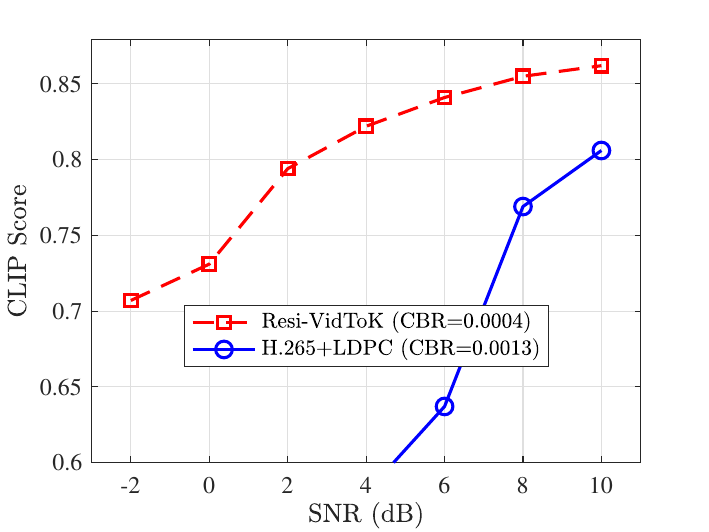}
    }
    \hfill
    \subfigure[UVG - CLIP Score]{
        \includegraphics[width=0.22\textwidth]{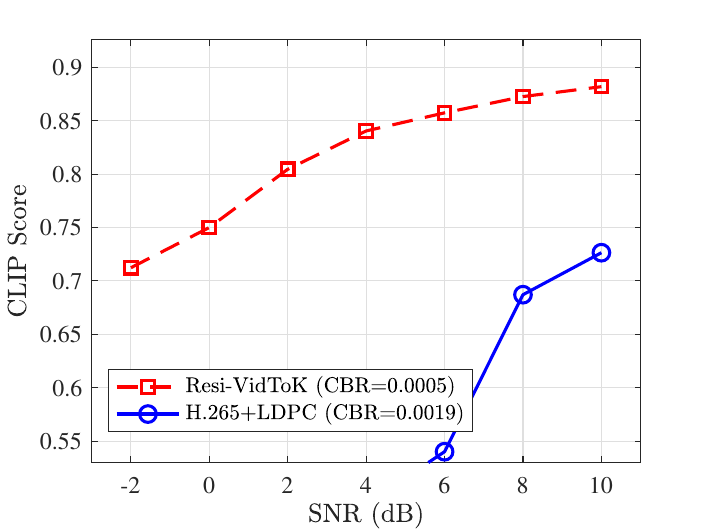}
    }
    \\ \vspace*{-3.9mm}
    
    \subfigure[WebVid - PSNR]{
        \includegraphics[width=0.22\textwidth]{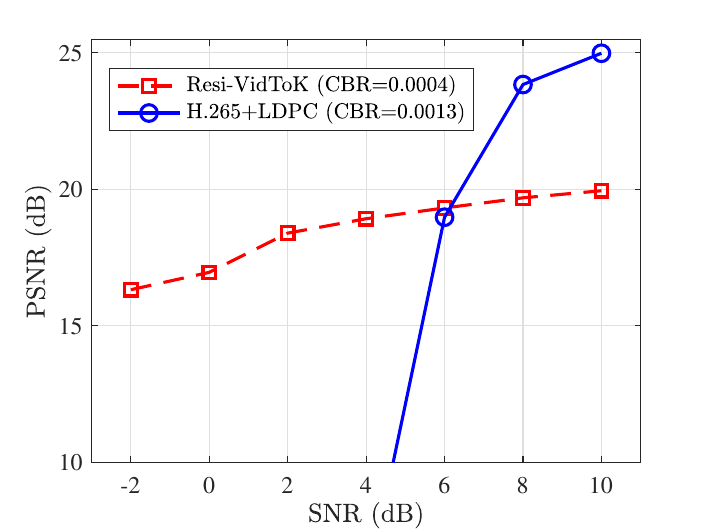}
    }
    \hfill
    \subfigure[UVG - PSNR]{
        \includegraphics[width=0.22\textwidth]{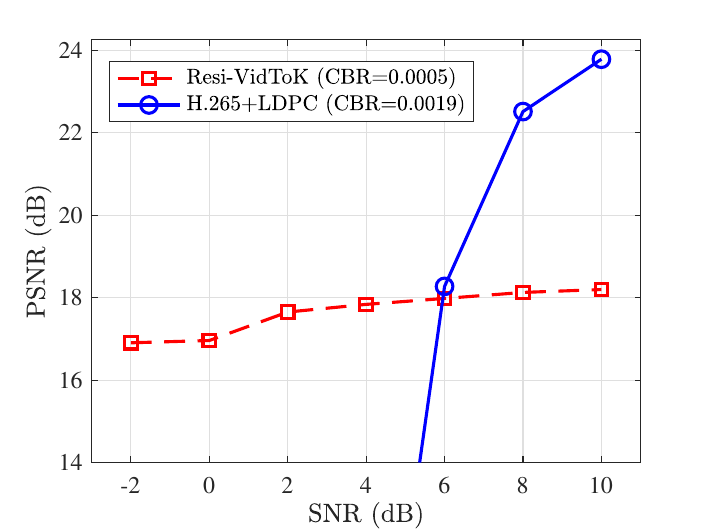}
    }
    \\ \vspace*{-3.9mm}
    
    \subfigure[WebVid - LPIPS]{
        \includegraphics[width=0.22\textwidth]{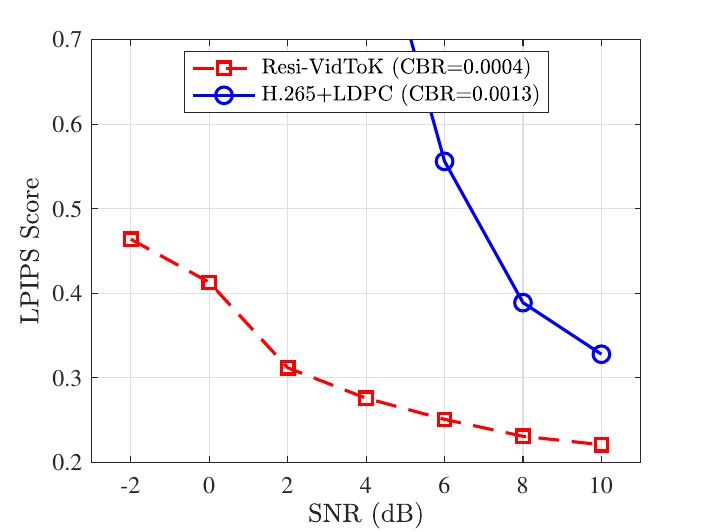}
    }
    \hfill
    \subfigure[UVG - LPIPS]{
        \includegraphics[width=0.22\textwidth]{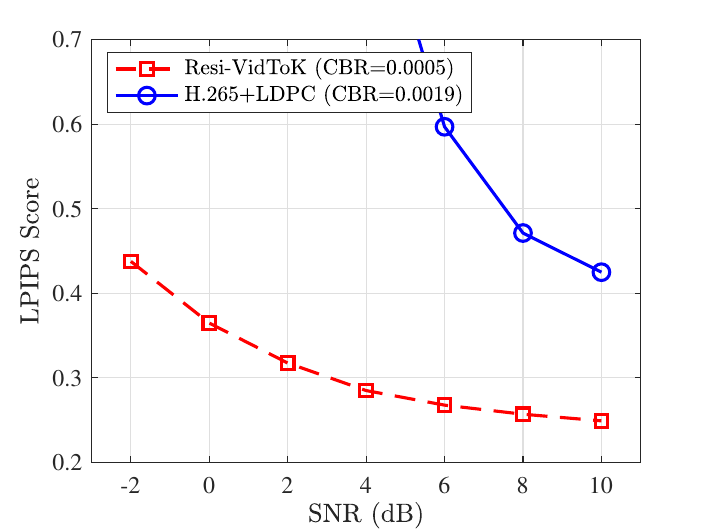}
    }
    
    \caption{Quality versus SNR with adaptive coding and modulation.}
    \vspace*{-3mm}
    \label{fig:snr_comparison}
\end{figure}

\subsection{Performance Comparison under Different SNR}
\label{subsec:snr_comparison}

Fig.~\ref{fig:snr_comparison} evaluates Resi\mbox{-}VidTok against the separated pipeline ``H.265\,+\,LDPC'' under \emph{ultra\mbox{-}low} rate budgets while sweeping the SNR from $-2$\,dB to $10$\,dB with adaptive coding and modulation (ACM; BLER target $0.002$; Table~\ref{tab:acm}). For WebVid we fix Resi\mbox{-}VidTok at $\mathrm{CBR}=4\!\times\!10^{-4}$, and for UVG at $\mathrm{CBR}=5\!\times\!10^{-4}$. Because H.265 at such extreme rates produces severe artifacts and frequent decode failures, the baseline uses a \emph{higher} rate—$1.3\!\times\!10^{-3}$ on WebVid and $1.9\!\times\!10^{-3}$ on UVG—i.e., more than $3\times$ the CBR of Resi\mbox{-}VidTok.

\paragraph{Semantic fidelity (CLIP).}
Across both datasets (Figs.~\ref{fig:snr_comparison}a–b), Resi\mbox{-}VidTok delivers consistently higher CLIP scores than H.265\,+\,LDPC \emph{at all tested SNRs}, despite operating at $<\!1/3$ the CBR. The CLIP curves for Resi\mbox{-}VidTok rise smoothly with SNR and quickly enter a high\mbox{-}quality regime, whereas the baseline requires both larger SNR and much higher rate to approach comparable semantics. This reflects the channel\mbox{-}adaptive top\mbox{-}$K$ strategy: MCS\mbox{-}first budgeting fixes the reliable bit count and the source coder spends those bits on the most informative tokens, protecting semantics even when the channel is weak.

\paragraph{Perceptual quality (LPIPS).}
Resi\mbox{-}VidTok achieves markedly lower (better) LPIPS than H.265\,+\,LDPC over the \mbox{$-2$–$10$\,dB} range on both WebVid and UVG (Figs.~\ref{fig:snr_comparison}e–f). The gap is largest at low SNR, where the separated pipeline struggles despite using $3\times$ higher CBR. Our design maintains perceptual realism by delivering a stable prefix of high\mbox{-}impact tokens and relying on efficient decoder\mbox{-}side interpolation to synthesize intermediate content without extra channel bits.

\paragraph{Structure (PSNR).}
For PSNR (Figs.~\ref{fig:snr_comparison}c–d), Resi\mbox{-}VidTok is competitive at low SNR and improves monotonically as SNR increases. At higher SNRs, the larger bit reservoir of H.265\,+\,LDPC can yield higher PSNR, as expected for an MSE\mbox{-}focused codec operating at $>\!3\times$ the rate. Nevertheless, in these same settings Resi\mbox{-}VidTok retains superior CLIP and LPIPS, indicating stronger semantic and perceptual fidelity per bit under extreme constraints.

\begin{figure}[t]
    \centering
    \includegraphics[width=0.5\textwidth]{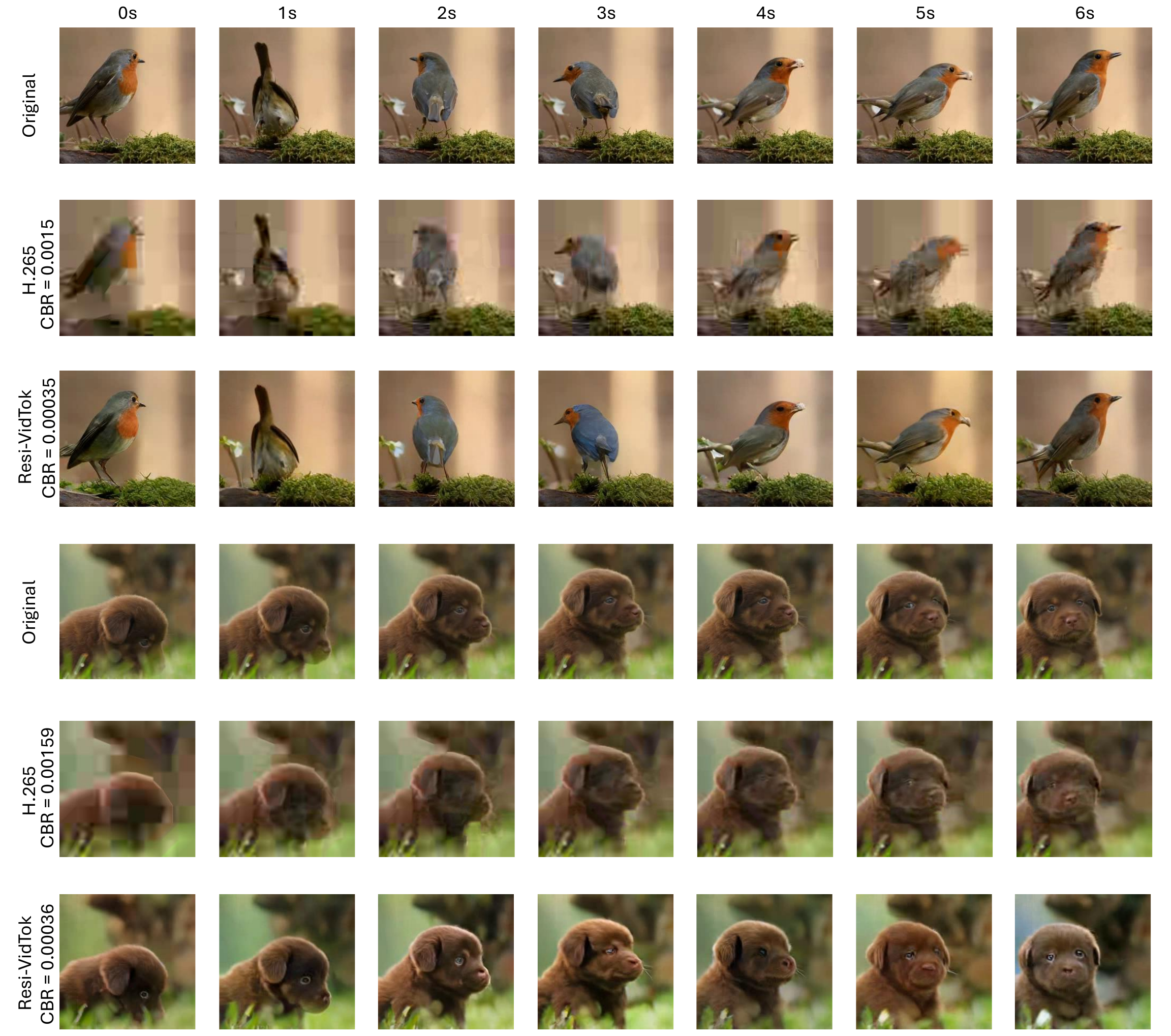}
    \caption{Visual comparison at SNR=6 dB and 16-QAM.}
    \label{fig:visual_demo}
    \end{figure}

    \subsection{Visualization}
    \label{subsec:visualization}
    
    Fig.~\ref{fig:visual_demo} qualitatively compares Resi\mbox{-}VidTok with the separated pipeline H.265\,+\,LDPC under challenging channel conditions ($\mathrm{SNR}=6$\,dB, 16\mbox{-}QAM). We show two examples (columns sampled every second from $0$\,s to $6$\,s): a \emph{bird} clip where Resi\mbox{-}VidTok operates at $\mathrm{CBR}=0.00035$ while H.265\,+\,LDPC uses $\mathrm{CBR}=0.00150$, and a \emph{dog} clip where Resi\mbox{-}VidTok uses $\mathrm{CBR}=0.00036$ versus $\mathrm{CBR}=0.00159$ for H.265\,+\,LDPC. Thus, the baseline consumes more than $3\times$ the bandwidth in both cases.
    
    \paragraph{Observations.}
    \emph{(i) Perceptual sharpness and semantics.} Resi\mbox{-}VidTok preserves fine structures and semantic cues—e.g., the bird’s eye and beak edges, feather contours, and the dog’s fur texture and eye highlights—despite operating at $\ll 10^{-3}$ CBR. In contrast, H.265\,+\,LDPC exhibits strong over\mbox{-}smoothing and “mushy” textures: edges are blurred, high\mbox{-}frequency details are suppressed, and object parts (wings, paws, eyes) appear less distinct, even though it uses $>3\times$ more bits. This gap aligns with our quantitative results where CLIP (semantic similarity) and LPIPS (perceptual similarity) favor Resi\mbox{-}VidTok at ultra\mbox{-}low rates.
    
    \emph{(ii) Temporal coherence.} Across time (0–6\,s), Resi\mbox{-}VidTok maintains stable appearance and motion continuity. Key frames provide strong anchors (thanks to top\mbox{-}$K$ token delivery), and the decoder\mbox{-}side interpolation reconstructs in\mbox{-}between frames without introducing block drift or flicker. The baseline often shows temporal inconsistency in flat regions and edges, a typical artifact of aggressive transform quantization at very low bitrate.
    
    \emph{(iii) PSNR vs.\ visual quality.} In these examples, the baseline can report higher PSNR at its much higher CBR, yet the visual results look blurrier. This underscores a known limitation of PSNR: it is MSE\mbox{-}oriented and favors over\mbox{-}smoothing at extreme compression, while human perception and downstream semantic measures (CLIP) reward the preservation of salient structures and textures. Resi\mbox{-}VidTok’s importance\mbox{-}ordered tokens and channel\mbox{-}adaptive top\mbox{-}$K$ strategy prioritize the bits that most improve semantics and perceptual realism, explaining the sharper, more faithful reconstructions.
    
 In summary, even when H.265\,+\,LDPC operates at more than triple the CBR, Resi\mbox{-}VidTok delivers crisper edges, richer textures, and more stable temporal appearance under the same channel and modulation settings. The visual comparison corroborates our metric trends (higher CLIP, lower LPIPS at ultra\mbox{-}low rate) and illustrates why a PSNR advantage for the baseline in some regimes does \emph{not} imply better perceived quality.
    
\subsection{Runtime on RTX A6000}
\label{subsec:runtime}

We benchmark the end\mbox{-}to\mbox{-}end Resi\mbox{-}VidTok pipeline on a single NVIDIA RTX A6000 GPU (input frame size $256{\times}256$) and observe an average throughput of $\sim$31 fps when stride $= 4$. The measurement includes all components for Resi-VidTok: tokenizer $f_{\mathrm{tok}}$, rate\mbox{-}adaptive top\mbox{-}$K$ header/body packing, channel coding/modulation, demodulation/decoding, reception and detokenization, shared framewise decoder $f_{\mathrm{dec}}$, and decoder\mbox{-}side RIFE interpolation.

\section{Conclusion}
\label{sec:conclusion}

We presented Resi\mbox{-}VidTok, a resilient, low\mbox{-}complexity framework for ultra\mbox{-}low\mbox{-}rate wireless video. Operating entirely in a discrete token space, it combines a shared framewise tokenizer with binary differential token compression, an efficient rate\mbox{-}adaptive source code that exploits a flexible header-body structure and transmits values only where changes occur, and an MCS\mbox{-}first channel adapter that converts instantaneous PHY capability into deliverable bits. Together with stride\mbox{-}controlled sampling and lightweight frame interpolation, Resi\mbox{-}VidTok achieves fast video recovery, maintains strong semantic and visual consistency at ultra-low bit budgets, and remains robust across a range of SNRs and CBRs



\bibliographystyle{IEEEtran}
\bibliography{SemRef}

\end{document}